\documentclass[journal]{IEEEtran}

\ifCLASSINFOpdf
\else
   \usepackage[dvips]{graphicx}
\fi
\usepackage{url}
\usepackage[acronym]{glossaries}
\usepackage{amssymb}
\usepackage{amsmath,amsfonts,amsthm}
\usepackage{multirow}
\usepackage{booktabs}
\usepackage{xcolor}
\usepackage[numbers,sort&compress]{natbib}

\usepackage[colorlinks, linkcolor=blue]{hyperref}
\usepackage{graphicx}

\newacronym{vae}{VAE}{variational autoencoder}
\newacronym{gru}{GRU}{gated recurrent unit}
\newacronym{rnn}{RNN}{recurrent neural network}
\newacronym{cnn}{CNN}{convolutional neural network}
\newacronym{stft}{STFT}{short-time Fourier transform}

\begin{document}

\title{Complex Recurrent Variational Autoencoder with Application to Speech Enhancement}

\author{Yuying Xie, Thomas Arildsen, and Zheng-Hua Tan, \IEEEmembership{Senior Member, IEEE}
\thanks{
The work of Yuying Xie is supported by China Scholarship Council.}
\thanks{The authors are with Department of
Electronic Systems, The Technical Faculty of IT and Design, Aalborg University, Denmark (E-mails: yuxi@es.aau.dk, tari@its.aau.dk, and zt@es.aau.dk).}
}

\maketitle

\begin{abstract}

As an extension of variational autoencoder (VAE), complex \acrshort{vae} uses complex Gaussian distributions to model latent variables and data. 
This work proposes a complex recurrent \acrshort{vae} framework, specifically in which complex-valued recurrent neural network and L1 reconstruction loss are used.
Firstly, to account for the temporal property of speech signals, this work introduces complex-valued recurrent neural network in the complex \acrshort{vae} framework.
Besides, L1 loss is used as the reconstruction loss in this framework.
To exemplify the use of the complex generative model in speech processing, we choose speech enhancement as the specific application in this paper. 
Experiments are based on the TIMIT dataset.
The results show that the proposed method offers improvements on objective metrics in speech intelligibility and signal quality.
\end{abstract}


\IEEEpeerreviewmaketitle

\section{Introduction}

\IEEEPARstart{C}{omplex} neural networks formulate a possible way to take full advantage of complex representations~\cite{CVNN_book}. Trabelsi et al. proposed the essential components of complex neural networks, and applied in complex convolutional neural networks~\cite{begio}. Wolter and Yao proposed a basic complex \acrfull{rnn} definition and complex \acrfull{gru} model, which shows positive stability and convergence property~\cite{wolter2018complex}.

\Acrfull{stft} representation, extensively applied in speech processing, is naturally complex-valued. Since an early study~\cite{griffin_lim} showed that magnitude is more important and structured, typical deep learning methods in speech enhancement have focused on magnitude spectrogram~\cite{Mortern_log_mag_SE}. However, the importance of phase has been proved~\cite{paliwal2011importance, le5phase} and thus complex spectrogram processing has attracted attention~\cite{phase_recent, phase_recent_book}.
Based on different representations of complex spectrograms in polar form and rectangular form, some works utilize magnitude and phase spectrograms~\cite{zheng2018phase, nugraha2019deep}, but most works use real and imaginary spectrograms~\cite{wang2020complex,fu2017complex,ouyang2019fully,deliang_wang,hu2020dccrn}. Different strategies have been applied by dealing with real and imaginary spectrograms concatenatedly~\cite{wang2020complex, fu2017complex} or separately~\cite{ouyang2019fully}. 
Considering the inherent relationship between real and imaginary spectra, using complex neural networks may be a more reasonable way. There also exist some works using complex-valued neural networks in speech processing. For instance, \cite{deliang_wang} applied complex feed-forward neural network in speech enhancement. In~\cite{hu2020dccrn} the authors proposed a complex model by combining a complex convolutional neural network with a real-valued recurrent neural network in an auto\-encoder framework. Deep complex u-net was proposed in~\cite{ICLR2018cu-net}.

\Acrfull{vae}~\cite{KingmaD.P2014AVB}, one of the most popular frameworks in representation learning, has been applied widely in speech processing~\cite{voice_conversion, leglaive2020recurrent}. 
As an extension of \acrshort{vae} theory, the complex \acrfull{vae} was derived mathematically in~\cite{nakashika2020complex} by assuming both latent variables and data following complex Gaussian distributions.
Experimental results in~\cite{nakashika2020complex} are positive on clean spectrogram reconstruction. 
However, since~\cite{nakashika2020complex} uses complex \acrshort{vae} only for clean spectrogram reconstruction without any specific application, the generalization of the model for speech signal has not been fully proved. 
Also, only linear layers are used in~\cite{nakashika2020complex} which significantly limits the expressive power of complex \acrshort{vae}.

Meanwhile, different objective functions have been applied in dealing with complex spectrograms for speech enhancement. Many works design the objective function in the complex spectrograms domain~\cite{Tan_ke_l2_loss} or in the (real) time domain~\cite{ICLR2018cu-net, hu2020dccrn}. Even though these loss functions can get higher scores on time-domain objective metrics, they may perform worse on speech intelligibility or quality evaluations.
Wang et al.~\cite{Jonathan_loss} pointed out that it is not sufficient for the objective function to only contain the time or complex spectrograms domain loss, but magnitude loss is also needed. Experimental results in~\cite{ICASSP2022_attention_L1_loss} also support this.


In this work, we propose a complex recurrent \acrshort{vae}. The contributions of this work mainly consist of three parts. 
First, as the Laplacian distribution can model speech spectrograms better~\cite{martin2003speech, ren-etal-2022-revisiting, laplace_mag, Laplace_magnitude_AAU}, we change the reconstruction loss function to L1 loss. Besides, based on the experiment results shown in~\cite{Jonathan_loss}, spectrogram magnitude loss and complex domain loss are all included in our objective function. 
Second, we introduce complex-valued GRU layers into the \acrshort{vae} framework to enable the model to utilise temporal information in speech signal processing. 
Third, the proposed method is demonstrated in speech enhancement. 



\section{Complex neural network}
\label{sec:cplx_neur_net}



\subsection{Complex feed-forward neural network}
For a complex-valued feed-forward neural network, if we use $\mathbf{W}=\mathfrak{R}(\mathbf{W})+i\mathfrak{I}(\mathbf{W})$ to denote complex weight, $\mathbf{x}=\mathfrak{R}(\mathbf{x})+i\mathfrak{I}(\mathbf{x})$ to denote input, and $\mathbf{b}=\mathfrak{R}(\mathbf{b})+i\mathfrak{I}(\mathbf{b})$ to denote bias, then the output of a complex-valued dense layer is:
\begin{equation}
\begin{aligned}
\mathbf{o} &= \mathbf{W}\mathbf{x} + \mathbf{b} \\
  &= [\Re(\mathbf{W})\Re(\mathbf{x})-\Im(\mathbf{W})\Im(\mathbf{x})+\Re(\mathbf{b})] \\&+ \imath[\Im(\mathbf{W})\Re(\mathbf{x})+\Re(\mathbf{W})\Im(\mathbf{x})+\Im(\mathbf{b})]\\
  &= \Re(\mathbf{o}) + \imath\Im(\mathbf{o})
\end{aligned}
\label{complex dense}
\end{equation}

\subsection{Complex activation functions}
Since complex values have different representations in rectangular form and polar form, complex-valued activation functions have many versions. The first proposed and mostly used complex-valued ReLU is modReLU~\cite{begio}, as shown in~\eqref{modRelu}:

\begin{equation}
\begin{aligned}
f_{\mathrm{modReLU}}(z) &= \mathrm{ReLU}(|z|+b)e^{i\theta_z} \\
&= \mathrm{ReLU}(|z|+b)\frac{z}{|z|}
\end{aligned}
\label{modRelu}
\end{equation}
where $z \in \mathbb{C}$, $|z|$ and $\theta _z$ are magnitude and phase of $z$, $b \in \mathbb{R}$ is a learnable parameter. 

Besides, a complex-valued sigmoid function, called modSigmoid, was proposed in~\cite{wolter2018complex} as shown in~\eqref{modsigmoid}, in which $\sigma(\cdot)$ denotes real-valued sigmoid function:
\begin{equation}
\begin{aligned}
f_{\mathrm{modSigmoid}}(z) = \sigma(\alpha \mathfrak{R}(z) + (1-\alpha) \mathfrak{I}(z)) \quad \alpha \in [0,1]
\end{aligned}
\label{modsigmoid}
\end{equation}
$\alpha$ used in this work equals 0.5.

\subsection{Complex recurrent neural network}
\label{sec:cplx_rec_nn}

The definition of basic complex \acrshort{rnn} is:
\begin{equation}
    \begin{aligned}
    \mathbf{z}_t = \mathbf{W}\mathbf{h}_{t-1} + \mathbf{V}\mathbf{x}_t + \mathbf{b} 
    \end{aligned}
\end{equation}
\begin{equation}
    \begin{aligned}
    \mathbf{h}_t = f_a(\mathbf{z}_t)
    \end{aligned}
\end{equation}
where $\mathbf{x}_t \in \mathbb{C}^{n_x \times 1}$, $\mathbf{h}_t \in \mathbb{C}^{n_h \times 1}$ denote input and hidden unit vector at time $t$, respectively, in which $n_x$, $n_h$ represent the dimension of $\mathbf{x}_t$ and $\mathbf{h}_t$. $\mathbf{W} \in \mathbb{C}^{n_h \times n_h}$, $\mathbf{V} \in \mathbb{C}^{n_h \times n_x}$ are hidden and input state transition matrices, respectively, while bias $\mathbf{b} \in \mathbb{C}^{n_h \times 1}$. $f_a(\cdot)$ is the point-wise nonlinear activation function.

Based on this, the complex GRU model is presented in the form of~\eqref{current memory content}-\eqref{final memory}:
\begin{equation}
\begin{aligned}
\tilde{\mathbf{z}}_t = \mathbf{W}(\mathbf{g}_r\odot \mathbf{h}_{t-1}) + \mathbf{V} \mathbf{x}_t + \mathbf{b}
\end{aligned}
\label{current memory content}
\end{equation}
\begin{equation}
    \begin{aligned}
    \mathbf{h}_t = \mathbf{g}_z \odot f_a(\tilde{\mathbf{z}}_t) + (1-\mathbf{g}_z)\odot \mathbf{h}_{t-1}
    \end{aligned}
    \label{final memory}
\end{equation}
$\odot$ represents Hadamard product. According to the experiment results and analysis in~\cite{wolter2018complex} and~\cite{unitary-bengio}, for stability, $f_a(\cdot)$ in~\eqref{final memory} is modReLU, and state transition matrices are unitary in this work. $\mathbf{g}_r$ and $\mathbf{g}_z$ represent reset gate and update gate, respectively, as shown in~\eqref{reset gate}-\eqref{update gate}:
\begin{align}
    \mathbf{g}_r &=f_g(\mathbf{z}_r), & 
    \mathbf{z}_r &= \mathbf{W}_r \mathbf{h}_{t-1} + \mathbf{V}_r \mathbf{x}_t + \mathbf{b}_r
    \label{reset gate}\\
    \mathbf{g}_z &= f_g(\mathbf{z}_z), &
    \mathbf{z}_z &= \mathbf{W}_z \mathbf{h}_{t-1}+ \mathbf{V}_z \mathbf{x}_t + \mathbf{b}_z
    \label{update gate}
\end{align}
where $f_g(\cdot)$ represents the modSigmoid function in~\eqref{modsigmoid}, $\mathbf{W}_r$, $\mathbf{W}_z \in \mathbb{C}^{n_h \times n_h}$ are state-to-state transition matrices,  $\mathbf{V}_r$, $\mathbf{V}_z\in \mathbb{C}^{n_h \times n_x}$ are input-to-state transition matrices, and $\mathbf{b}_r$, $\mathbf{b}_z \in \mathbb{C}^{n_h}$ are biases.

\subsection{Initialization}
\label{sec:initialization}

In this paper, the initialization of weights in both complex dense layers and complex \acrshort{gru} layers follows~\cite{wolter2018complex} and~\cite{glorot_initializer}, i.e., weights are sampled from uniform distribution $\mathcal{U}$[-$A$, $A$], where $A=\sqrt{6 / (n_{in}+n_{out})}$. $n_{in}$ and $n_{out}$ are the dimensions of input and output. All biases are initialized as 0, except for $\mathbf{b}_r$ and $\mathbf{b}_z$ in \acrshort{gru} layers, for which 4 is used as initialization value as in~\cite{wolter2018complex}.

\subsection{Gradient calculation}

Assume function $f(z)$ is real-valued with complex variable $z$, i.e., $f:\mathbb{C}\mapsto\mathbb{R}$, then $f(z)$ is non-holomorphic as long as $f(z)\not\equiv0$. Presume $z=x+\imath y$, $x,y\in \mathbb{R}$.  Wirtinger calculus is used for partial derivative of a non-holomorphic function:
\begin{align}
        \frac{\partial f}{\partial z} &= \frac{1}{2} \left(\frac{\partial f}{\partial x} - \imath \frac{\partial f}{\partial y}\right) \label{wirtinger calcus1} \\
        \frac{\partial f}{\partial z^*} &= \frac{1}{2} \left(\frac{\partial f}{\partial x} + \imath \frac{\partial f}{\partial y}\right)
    \label{wirtinger calcus2}
\end{align}

\section{Complex recurrent variational autoencoder}

\label{sec:cplx_rec_vae}
\subsection{Complex variational autoencoder}
In the complex-valued \acrshort{vae} framework~\cite{nakashika2020complex}, $\mathbf{x} \in \mathbb{C}^{n_x}$, $\mathbf{z} \in \mathbb{C}^{n_z}$ are used to represent data and latent variable, respectively, in which $n_x$ and $n_z$ denote dimensions of data and latent variable. As with conventional \acrshort{vae}, the loss function of the complex-valued \acrshort{vae} is:
\begin{equation}
    \begin{aligned}
    \log p(\mathbf{x}) &\ge \mathcal{L}(\theta, \phi;\mathbf{x}) \\
    &= \mathcal{L}_{rec}(\mathbf{x}) - KL(q_{\phi}(\mathbf{z}|\mathbf{x})||p(\mathbf{z})) \\
    &= \mathbb{E}_{q_{\phi}(\mathbf{z}|\mathbf{x})}[\log p_{\theta}(\mathbf{x}|\mathbf{z})] - KL(q_{\phi}(\mathbf{z}|\mathbf{x})||p(\mathbf{z}))
    \end{aligned}
    \label{ELBO}
\end{equation}
$\phi$ and $\theta$ denote parameters in inference model and generative model individually, and $p$ and $q$ represent, respectively, the prior and posterior distributions.

\begin{figure*}[htb]
    \centering
    \includegraphics[width=12.0cm]{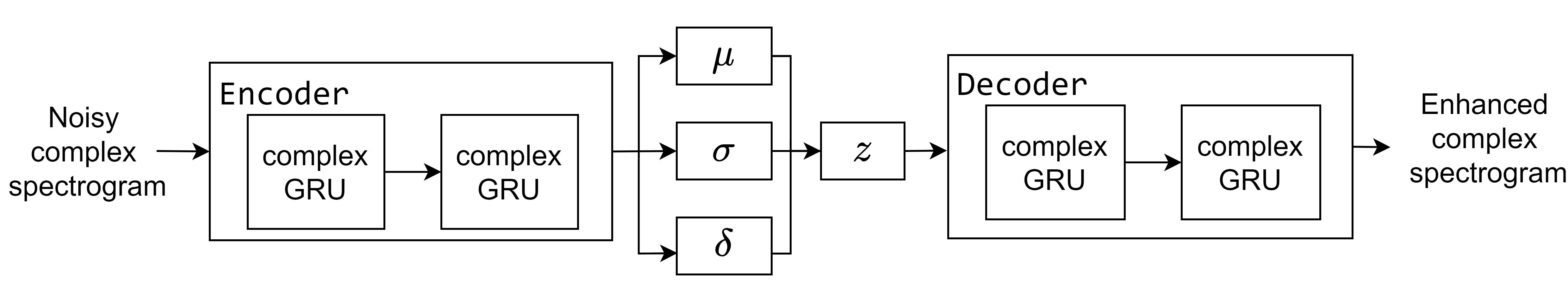}
    \caption{Structure of complex recurrent VAE}
    \label{fig:COMPLEXvae}
\end{figure*}

Compared to real-valued \acrshort{vae}, complex \acrshort{vae} not only uses complex-valued parameters in the whole framework, but it also changes the assumption of data space and latent variable space distributions.
A multivariate complex normal distribution is used to model latent variable and data in complex-valued \acrshort{vae}. If a complex random variable $\mathbf{h} \in \mathbb{C}^D$ follows a multivariate complex normal distribution, i.e., $\mathbf{h} \sim \mathcal{N}_c(\mathbf{a},\mathbf{\Gamma}, \mathbf{C})$, in which $\mathbf{a} \in \mathbb{C}^D$, $\mathbf{\Gamma} \in \mathbb{C}^{D \times D}$, $\mathbf{C}  \in \mathbb{C}^{D \times D}$ denote mean vector, covariance matrix and pseudo-covariance matrix in order, then the probability density of $\mathbf{h}$ can be written as:
\begin{equation}
    \begin{aligned}
    p(\mathbf{h})  & \triangleq\frac{1}{\pi ^D \sqrt{\det(\mathbf{\Gamma}) \det(\Bar{\mathbf{\Gamma}}-\mathbf{C}^H \mathbf{\Gamma} ^{-1} \mathbf{C})}} \\
    & \cdot \exp\left\{{-\frac{1}{2}}\left[\begin{array}{cc}
         \mathbf{h}-\mathbf{a} \\ \mathbf{\Bar{h}}-\mathbf{\Bar{a}} 
    \end{array}\right]^{H}
    \left[ \begin{array}{cc}
         \mathbf{\Gamma} & \mathbf{C} \\
         \mathbf{C}^H & \mathbf{\Gamma}^H 
    \end{array} \right]^{-1}
    \left[ \begin{array}{cc}
         \mathbf{h}-\mathbf{a} \\ \mathbf{\Bar{h}}-\mathbf{\Bar{a}}
    \end{array} \right]
    \right\}
    \end{aligned}
    \label{complex normal distribution}
\end{equation}

For data $\mathbf{x}$, if we assume its prior distribution $p_{\theta}(\mathbf{x}|\mathbf{z}) = \mathcal{N}_c(\mathbf{x};\mathbf{a},\mathbf{I},\mathbf{0})$, according to \eqref{complex normal distribution}, we get:
\begin{equation}
    \begin{aligned}
    \mathbb{E}_{q_{\phi}(\mathbf{z}|\mathbf{x})}[\log p_{\theta}(\mathbf{x}|\mathbf{z})] \approx - \| \mathbf{x}-\mathbf{a} \| _2 ^2 + K
    \end{aligned}
    \label{decoder loss}
\end{equation}
in which $K$ is a constant.

For latent variable $\mathbf{z}$, we assume its posterior follows a multivariate complex normal distribution with diagonal covariance and pseudo-covariance matrices, i.e., $q_{\phi}(\mathbf{z}|\mathbf{x})=\mathcal{N}_c(\mathbf{z}; \boldsymbol{\mu}, \Delta(\boldsymbol{\sigma}), \Delta(\boldsymbol{\delta}))$, in which $\boldsymbol{\mu} \in \mathbb{C}^{n_z}$, $\boldsymbol{\sigma} \in \mathbb{R}_+^{n_z}$, $\boldsymbol{\delta} \in \mathbb{C}^{n_z}$, and $\Delta(\cdot)$ represents diagonal matrix. Assume the prior of $\mathbf{z}$ follows a standard complex normal distribution, i.e., $p(\mathbf{z}) \triangleq \mathcal{N}_c(\mathbf{0},\mathbf{I},\mathbf{0})$. Then, the KL divergence in~\eqref{ELBO} can be written as:
\begin{equation}
    \begin{aligned}
    &KL(q_{\phi
    }(\mathbf{z}|\mathbf{x})||p(\mathbf{z})) \\
    &= KL(\mathcal{N}_c(\boldsymbol{\mu}, \Delta(\boldsymbol{\sigma}), \Delta(\boldsymbol{\delta}))||\mathcal{N}_c(\mathbf{0},\mathbf{I},\mathbf{0}))\\
    &= \boldsymbol{\mu} ^H \boldsymbol{\mu} + \| \boldsymbol{\sigma} -\mathbf{1} -\frac{1}{2} \log (\boldsymbol{\sigma} ^2 - |\boldsymbol{\delta}|^2) \| _1
    \end{aligned}
\end{equation}

In real-valued \acrshort{vae}, the 'reparameterization trick' introduced in~\cite{KingmaD.P2014AVB} has been used to make back-propagation achievable from decoder to encoder. A similar trick is also needed in complex \acrshort{vae}. Under the assumption of the estimated latent variable $\mathbf{\tilde{z}} \sim \mathcal{N}_c (\mathbf{z}; \boldsymbol{\mu}, \Delta(\boldsymbol{\sigma}), \Delta(\boldsymbol{\delta}))$, and the elements of latent variable $\mathbf{z}$ are independent of each other, then 
\begin{equation}
    \begin{aligned}
    \mathbf{\tilde{z}} = \boldsymbol{\mu} + \mathbf{k}_r \odot \boldsymbol{\epsilon}_r + \mathbf{k}_i \odot \boldsymbol{\epsilon}_i
    \end{aligned}
    \label{reparameterization of complex Gaussian}
\end{equation}
\begin{equation}
\begin{aligned}
\mathbf{k}_r = \frac{\boldsymbol{\sigma} + \boldsymbol{\delta}}{\sqrt{2\boldsymbol{\sigma} + 2\Re(\boldsymbol{\delta})}}
\end{aligned}
\label{kr}
\end{equation}
\begin{equation}
    \begin{aligned}
    \mathbf{k}_i = \imath \frac{\sqrt{\boldsymbol{\sigma} ^2 - |\boldsymbol{\delta}|^2}}{\sqrt{2\boldsymbol{\sigma} + 2\Re(\boldsymbol{\delta})}}
    \end{aligned}
    \label{ki}
\end{equation}
$\boldsymbol{\epsilon}_r$ and $\boldsymbol{\epsilon}_i$ are random variables following a standard Gaussian distribution, i.e., $\boldsymbol{\epsilon}_r, \boldsymbol{\epsilon}_i \sim \mathcal{N}(\mathbf{0}, \mathbf{I})$. And $\sqrt{\cdot}$ in~\eqref{kr}-\eqref{ki} means element-wise square root.

\subsection{Proposed method}



%

Even though the real-valued Gaussian distribution has been widely used in \acrlong{vae} architecture, it causes blurry reconstruction due to its tolerance of small deviation~\cite{gaussian_disadvantange1}. The same blurry reconstruction phenomena also occur in the complex-valued domain. Besides, from the aspect of improving speech enhancement performance on evaluation metrics, paper~\cite{Jonathan_loss} shows that when dealing with complex-valued spectrograms, it is not enough to only consider time or complex spectrograms domain loss, but also the magnitude loss to perform betters on speech quality and intelligibility. 
Experimental results in paper~\cite{Jonathan_loss, example_of_l1_loss} also confirm this finding.
Therefore, inspired by paper~\cite{Jonathan_loss}, the reconstruction loss we used in this work is:
\begin{equation}
    \begin{aligned}
     \mathcal{L}_{rec}
    = 
    \| \Re(\mathbf{x}) - \Re(\mathbf{\hat{x}}) \|_1
    + \| \Im(\mathbf{x}) - \Im(\mathbf{\hat{x}}) \|_1 
    + \| |\mathbf{x}| - |\mathbf{\hat{x}}| \| _1
    \end{aligned}
    \label{proposed loss}
\end{equation}
in which $\mathbf{x}$ and $\mathbf{\hat x}$ are the truth value and the neural network output.
As maximizing likelihood assuming a Laplacian distribution is equivalent to minimizing mean absolute error, \eqref{proposed loss} can be regarded as the real, imaginary and magnitude spectrograms assuming to follow Laplacian distribution with unit scale.

Compared to the Gaussian distribution, the Laplacian distribution exhibits a sharper peak and leads to more precise reconstruction results~\cite{gaussian_disadvantange1}. 
Paper~\cite{martin_2002_laplacian_histogram_complex} illustrates that, for short DFT (frame size $<$ 100ms), Laplacian distribution fits real and imaginary parts of speech coefficients better than the Gaussian distribution, and experimental results in~\cite{martin2003speech} prove this.
Besides, paper~\cite{laplace_mag} illustrate that Laplacian distribution fits magnitude spectrograms well based on moment test results, and has supported loss design in paper~\cite{, ren-etal-2022-revisiting} with positive results.
Thus, it is reasonable to use eq.~\eqref{proposed loss} as the reconstruction loss function. Positive results from other works also prove effectiveness of eq.~\eqref{proposed loss}~\cite{example_of_l1_loss}.

Complex \acrshort{vae} with linear layers for direct representation learning of complex spectrograms was proposed in~\cite{nakashika2020complex}. Besides changing the reconstruction loss, since the recurrent neural network has a natural advantage in time-series signal processing, we propose a complex recurrent \acrshort{vae} by using complex-valued \acrshort{gru} in a complex \acrshort{vae} framework. Figure~\ref{fig:COMPLEXvae} shows the structure of the complex recurrent \acrshort{vae}. Both encoder and decoder are composed of two complex-valued \acrshort{gru} layers as mentioned in Section~\ref{sec:cplx_rec_nn}. The encoder outputs parameters of the posterior of the latent variable $\mathbf{z}$, i.e., mean $\boldsymbol{\mu}$, covariance $\boldsymbol{\sigma}$ and pseudo-covariance $\boldsymbol{\delta}$. The latent variable $\mathbf{z}$ resulting from the re-parameterization trick is fed into the decoder. As an exploration of using a complex generative model in speech processing, speech enhancement is chosen to test the performance.

\section{Experiment}
\label{sec:experiment}

\begin{table*}
  \renewcommand\arraystretch{1.2}
  \caption{Performance of different models for speech enhancement}
  \label{tab:results}
  \centering
  \scriptsize
  \begin{tabular}{ c|c|c|c|c|c|c|c|c|c|c|c|c|c}
    \hline
    \multicolumn{2}{@{}c@{}|}{\textbf{Noise type}} & \multicolumn{4}{@{}c@{}|}{\textbf{ESTOI}}
    & \multicolumn{4}{@{}c@{}|}{\textbf{SI-SDR}(dB)} & \multicolumn{4}{@{}c@{}}{\textbf{PESQ}} \\ \hline

    \multicolumn{2}{@{}c@{}|}{} & noisy & R-RVAE & C-VAE & C-RVAE & noisy & R-RVAE & C-VAE & C-RVAE & noisy & R-RVAE & C-VAE & C-RVAE \\
    \hline
    
    \multirow{5}{*}{Seen} &  BBL & 0.44 & 0.47$\pm$0.00 & 0.52$\pm$0.05 & \textbf{0.55$\pm$0.00} & 
                                        1.05 & 2.40$\pm$0.24 & 4.64$\pm$2.10 & \textbf{5.69$\pm$0.07} &
                                        1.72 & 1.77$\pm$0.02 & 2.00$\pm$0.03 & 2.00$\pm$0.05\\
     
    \cline{2-14}

     &  CAF & 0.53 & 0.61$\pm$0.01 & 0.60$\pm$0.06 & \textbf{0.67$\pm$0.02} & 
              -0.05 & 5.36$\pm$0.11 & 6.91$\pm$2.56 & \textbf{8.95$\pm$0.89} &
              1.73 & 2.16$\pm$0.03 & 2.20$\pm$0.06 & \textbf{2.35$\pm$0.05}\\
     
    \cline{2-14}

    &  SSN & 0.46 & 0.49$\pm$0.01 & 0.50$\pm$0.06 & \textbf{0.56$\pm$0.02} & 
             0.69 & 2.58$\pm$0.72 & 3.55$\pm$2.54 & \textbf{5.37$\pm$0.42} &
             1.45 & 1.81$\pm$0.04 & 1.84$\pm$0.07 & \textbf{1.96$\pm$0.02}\\
     
    \cline{2-14}

    &  STR & 0.54 & 0.63$\pm$0.01 & 0.65$\pm$0.07 & \textbf{0.73$\pm$0.02} & 
            2.27 & 6.13$\pm$0.19 & 8.39$\pm$2.69 & \textbf{10.66$\pm$1.09} &
            1.79 & 2.28$\pm$0.04 & 2.29$\pm$0.07 & \textbf{2.46$\pm$0.05}\\
     
    \cline{2-14}

    &  \footnotesize{\textbf{AVE}} & 0.49 & 0.55$\pm$0.01 &  0.57$\pm$0.06 & \textbf{0.63$\pm$0.01} & 
             0.99& 4.03$\pm$0.27 & 5.87$\pm$2.47 & \textbf{7.67$\pm$0.60} &
            1.67& 2.00$\pm$0.02 & 2.08$\pm$0.05 & \textbf{2.20$\pm$0.01}\\
     
    \cline{1-14}

    \multirow{3}{*}{Unseen} & BUS & 0.62 & 0.67$\pm$0.00 & 0.67$\pm$ 0.05& \textbf{0.73$\pm$0.02} & 
                                        2.09 & 6.82$\pm$0.30 & 8.36$\pm$2.74 & \textbf{10.37$\pm$1.10} &
                                        2.15 & 2.48$\pm$0.03 & 2.45$\pm$0.04& \textbf{2.51$\pm$0.02}\\
     
    \cline{2-14}

    & PED & 0.49 & 0.55$\pm$0.00 & 0.56$\pm$0.05& \textbf{0.63$\pm$0.02} & 
           1.92 & 4.06$\pm$0.11 & 5.04$\pm$2.54 & \textbf{6.91$\pm$0.69} &
           1.62 & 1.97$\pm$0.02 & 1.96$\pm$0.05& \textbf{2.13$\pm$0.07} \\
     
    \cline{2-14}

    & \footnotesize{\textbf{AVE}} & 0.56 & 0.62$\pm$0.00 & 0.62$\pm$0.05 & \textbf{0.68$\pm$0.02} & 
           2.01& 5.36$\pm$0.17 & 6.70$\pm$2.64 & \textbf{8.64$\pm$0.89} &
           1.89& 2.22$\pm$0.02 & 2.21$\pm$0.04 & \textbf{2.32$\pm$0.04}\\
     
    \cline{1-14}
  \end{tabular}
\end{table*}

\subsection{Dataset}
The TIMIT dataset~\cite{TIMIT} is used for the following experiments. 
All 4620 utterances, 500 utterances and 192 utterances in the TIMIT training, development and core test sets are used for training, validation and test separately. 
The noise dataset, as the same one used in paper~\cite{noise_dataset}, contains 6 different noise types: babble (BBL), cafeteria (CAF), street (STR), speech shaped noise (SSN), bus (BUS) and pedestrian (PED).
The first four noise types are used for training, development and test set generation as seen noise type, while the last two types are used as unseen noise type only in test set generation. To generate the noisy speech dataset, for each utterance in the training and development sets, SNR is selected uniformly at random from -10 dB to 10 dB with 1 dB as step size. For model evaluation, noisy speech data with SNR equal to $\{$-6, -3, 0, 3, 6$\}$ dB has been produced separately. Clean speech signals are all normalized to have unit RMS power each before adding noise, and only the speech-active region of clean speech signals has been used to calculate the scale of noise to attain the target SNR.
The sampling rate equals 16kHz. 200-dimensional complex spectrogram is used as input for the complex neural network, while the real-valued neural network utilizes the same dimension log-magnitude spectrogram.

\subsection{Baseline}
To compare a real-valued neural network to a complex-valued neural network, one baseline is a real-valued recurrent \acrshort{vae}, denoted R-RVAE. Both encoder and decoder of R-RVAE contain two GRU layers, like in Fig.~\ref{fig:COMPLEXvae}, but real-valued. The prior of the latent variable $\mathbf{z}$ is the standard Gaussian distribution, and the posterior of $\mathbf{z}$ is $\mathcal{N}(\boldsymbol{\mu}, \boldsymbol{\sigma}^2)$, in which $\boldsymbol{\mu}$ and $\boldsymbol{\sigma}$ are from neural network. The reconstruction loss used for R-RVAE is L2 loss. 

Inspired by work in~\cite{deliang_wang}, complex-valued feed-forward \acrlong{vae}, denoted C-VAE, has been chosen as another baseline. For a fair comparison, the structure of C-VAE contains four complex dense layers. ModReLU has been used as activation function after every layer except the encoder and decoder output layer. The reconstruction function of C-VAE is as~\eqref{proposed loss}.

\subsection{Implementation details}
The proposed complex recurrent \acrshort{vae} is denoted C-RVAE. For all models, the batch size equals 100, while the learning rate is $10^{-5}$. For C-RVAE, the \acrshort{gru} layers contain 512 units in both encoder and decoder, while the latent variable dimension equals 512. 
The inputs of C-RVAE and R-RVAE are 2 frames complex-valued or log-magnitude spectrogram.

To have the same parameter number, R-RVAE uses 939-unit \acrshort{gru} layers in encoder and decoder, and the latent variable dimension also equals 939. To make a fair comparison, the first three layers of C-VAE contain 512 units. For all models, training stops when the loss has not decreased for 50 epochs. We work in Tensorflow~\cite{abadi2016tensorflow}, and the gradient optimizer used here is Adam~\cite{kingma2014adam}. The weight initializations of complex neural networks all follow Section~\ref{sec:initialization}. \href{https://github.com/yuxi6842/CRVAE}{Code} is published.\footnote{https://github.com/yuxi6842/CRVAE}

\subsection{Results and discussion}

Results of speech enhancement have been evaluated by extended short-time objective intelligibility (ESTOI) measure~\cite{ESTOI} (range from 0 to 1), scale-invariant signal-to-distortion ratio (SI-SDR) measured in dB~\cite{SI-SDR} (range from $-\infty$ to $+\infty$), and perceptual evaluation of speech quality (PESQ) measure~\cite{PESQ} (range from 1 to 4.5) to evaluate speech intelligibility, signal quality, and speech quality, respectively. Higher score means better performance for all measures.

All models are broadly trained under 4 seen noise conditions and tested under all noise conditions. Table~\ref{tab:results} shows the results of the different models under conditions of seen noise types and unseen noise types, respectively. 
The number in every cell is the evaluation results averaged over scores from test sets with 5 different SNRs. Every experiment has been repeated 5 times to get mean and standard deviation results.
The best result in each row is highlighted in bold font. The unprocessed noisy speech has also been evaluated for comparison and been listed with title 'noisy', while the average scores over different noise types have been listed in the rows with 'AVE'.

 
Based on the results shown in Table~\ref{tab:results}, we can make the following conclusion:

1. C-RVAE shows positive results on ESTOI/SI-SDR, and slightly better scores on PESQ, compared to R-RVAE and C-VAE, under both known and unknown noise conditions. 

2. Application of a recurrent neural network in complex VAE improves enhanced speech intelligibility and signal quality, compared to C-VAE.

\section{Conclusions}
\label{sec:conclusions}

Motivated by the modeling power of complex \acrshort{vae} and the temporal property of speech signals, this work expands complex linear \acrshort{vae} to nonlinear and recurrent \acrshort{vae}. The proposed method has been applied in speech enhancement based on the TIMIT dataset. ESTOI, SI-SDR and PESQ are utilized as evaluation methods for speech intelligibility, signal fidelity, and speech quality, respectively. Experiments show that, compared to real-valued recurrent \acrshort{vae} and complex feed-forward variational autoencoder, complex recurrent \acrshort{vae} has obvious advantages in terms of signal fidelity, and shows positive results on speech quality and intelligibility. The experiments prove the modelling capability of complex recurrent VAE, and may pave an avenue for complex-valued latent representation learning.

\small
\bibliographystyle{IEEEtran}
\bibliography{bare_jrnl}
\end{document}